\documentclass[11pt,twoside]{book}
\usepackage{ml-books}
\usepackage{url,hyperref}
\usepackage{graphics,graphicx,epsfig}
\usepackage{amssymb,color}
\usepackage{amsmath}

\begin{document}

%\contribauth{William Bialek}
\chapter*{Optimization principles, revisited}
%\label{ch:first}
\contrib{William Bialek}
\contribnote{Joseph Henry Laboratories of Physics, and \\
Lewis-Sigler Institute for Integrative Genomics\\ 
Princeton University\\
Initiative for the Theoretical Sciences\\
The Graduate Center of the City University of New York}

\begin{quote}
The brain solves a wide range of problems, and some of these solutions are very good indeed.  It is an old idea that, in some cases, these solutions may be as good as possible given the relevant constraints.  If this is true we can  promote this notion of optimization to a principle from which aspects of function and mechanism can be {\em derived}, ideally with no free parameters.  Optimization principles are powerful and have had many successes, but they remain controversial. We start with a reminder of a classic example---diffraction limited optics in insect eyes---and then a summary of why you should worry about optimization as a more general principle.  Next we turn to recent progress on the theory of optimal motion estimation in fly vision, and finally summarize  recent results on the landscape for optimization, arguing that (contrary to widely held intuitions) optimal performance need not require the underlying mechanisms to have finely tuned parameters.  I hope to convince you that this is a moment to take optimization seriously as the basis for theories of neural coding and computation.  This is based on a rather informal lecture at the summer school LASCON X in S\~ao Paulo,  and I have tried to preserve that tone here; pointers are given to more detailed and technical discussions. 
\end{quote}

\section{Introduction}

In order to survive in the world organisms have to solve a wide range of problems.  We know a great deal about the microscopic components that make these solutions possible, but much less about how the parts fit together into a functional whole.  What we do know is that random collections of interacting parts typically are {\em not} functional.\footnote{This starts at the most microscopic level:  random sequences of amino acids do not fold into well defined structures that can carry out biochemical functions.}  Thus the systems that we see today---the amino acid sequences of proteins, the networks of interacting genes in a developing embryo, the connections among neurons in a brain---have been selected from the set of possible mechanisms to achieve some level of functional performance.  Might this selection be sufficiently strong that it has predictive power?  Could natural systems be the ones that maximize performance subject to some sensible constraints?  Even if organisms don't maximize performance, could a theory of optimal systems be a useful idealization?

More than three centuries ago, Leibniz gave us tools for finding the maxima and minima of functions.  He also articulated the idea that we live in the best of all possible worlds, taking maximization to an extreme \cite{leibniz}.  Leibniz' concerns were theological and not scientific, and his extraordinary optimism has been criticized and  parodied, repeatedly.  Perhaps the most famous parody is Voltaire's Professor Pangloss, who gives voice to Leibniz' claim in {\em Candide}, despite the accumulation of disasters around him \cite{voltaire}.  Not long after Darwin, du Bois--Reymond suggested that evolution might lead to essentially perfect adaptation, echoing at least the spirit of Leibniz' original claim \cite{dubois}.  A century later, Gould and Lewontin described this focus on adaptation as the ``Panglossian paradigm,'' and emphasized instead the roles of randomness, contingency, and constraint in shaping the organisms that we see today \cite{gould+lewontin_1979}.

On a dark night, the arrival of single photons at our retina makes a measurable contribution to our perception.\footnote{This is a subject with a beautiful history, starting not long after the quantum nature of light was discovered.  If we follow the problem from photon absorption by rhodopsin all the way to the perceptual decision that we have seen something, we touch many of crucial issues in the physics of life more generally.  See Chapter 2 of Ref \cite{bialek_2012}.} The reliability with which we can count these single photon events is limited by the dark noise in the rod photoreceptors, and this is possible only because the retinal circuitry implements filters and nonlinearities that are matched to the structure of signals and noise generated by these cells.  In a precise sense, it would be very hard for the system to do better: photon counting in vision provides a concrete example of optimization in biological function.  Is it Panglossian to think that optimization may be a more general principle?

In keeping with Leibniz, many view the optimization of biological function as a philosophical (or perhaps even theological) idea.  Some believe that optimization is obvious because evolution has had a billion years to get things right.  Others believe that it's impossible, not least because evolution in a dynamic environment is not guaranteed to reach an optimum.  My experience is that, for many people, these general views largely are unaffected by evidence for or against optimization in specific cases.  

More seriously, it is not clear what evolution has to say about the question of optimization.  In order to conclude, for example, that evolution favors the emergence of single photon counting we need to understand both the benefits and the costs.  The benefits seem obvious: a more sensitive visual system allows organisms to search for food and mates under a wider range of conditions.  But the costs are substantial, since the receptors that count single photons are among the most metabolically active cells in our body.  While we might be able to estimate this metabolic cost, we have no idea how to compare this with the benefits.  Absent such a comparison, it is not clear how we could even start to understand what evolutionary theory predicts.

Optimization, used well, is a quantitative theory.  Contrary to being a matter of belief, we can use the idea of optimization to discipline our thinking.  If we want to say that something is the best of all the possibilities, we need to be clear what we mean by best and what we consider to be possible.  If we  suggest that “biological systems maximize X” is a principle, then we are making explicit our intuition that X matters in the life of the organism, or more precisely that X measures performance at a function essential for life.  These intuitive claims could simply be wrong.  To make progress we will need a mathematically precise definition of X, and there may be many different formalizations of our intuition that we need to sort through.

If maximizing X is going to serve as a principle,  we have to understand why X can’t be arbitrarily large: we need  a theory that defines the relevant constraints and the resulting physical limits to performance. Ideally we should actually be able to measure X and compare its value with this theoretical maximum; often this suggests new experiments or new ways of looking at existing data. Finally, maximizing X should generate some definite predictions about the structure and dynamics of the system, predictions that can be tested in independent quantitative experiments.   

The idea of deriving quantitative predictions from general principles is of course familiar in physics, but in the context of living systems it often feels that the best we can do is to build models with many parameters and fit them to the data.   Optimization provides a way of cutting through this complexity, and in favorable cases results in quantitative predictions that have no free parameters \cite{bialek_2024}.  Replacing parameters with principles, allowing us to derive rather than fit the behavior that we see, is a grand ambition.  It would be wonderful if we succeeded, but it is a serious program precisely because it has the possibility of failure.

In one short lecture I could not possibly give a full survey of optimization in neural coding and computation.   Instead I sketched two examples, one classical and one modern, with the latter including work in progress.  Interleaved between these examples is some discussion of the reasons to be suspicious of optimality as a general principle, and then at the end I look at recent developments which  respond directly to these objections.  In writing these notes I have tried to keep the same informal tone of the original lecture, including a few personal notes.

\section{A classical example}

In the nineteenth century, exploration of vision and hearing was entwined with the emerging understanding of light and sound.  Major figures of classical physics---Helmholtz, Maxwell, Ohm, and Rayleigh---all moved freely between  subjects that for much of the twentieth century we might have distinguished as physics and neuroscience or psychology.  Their search for a physicist's understanding of perception provides, in many ways, a model for the modern version of biological physics as a branch of physics \cite{decadal}.

Not long after it was understood that diffraction sets a limit to the sharpness of images it was realized that this is relevant for our own eyes, and the limits would be even more severe in the compound eyes of insects with their many small lenses \cite{mallock_1894}.  Fifty years later Balow realized that these limitations lead naturally to an optimization problem \cite{barlow_1952}.

\begin{figure}[t]
\centerline{\includegraphics[width = \linewidth]{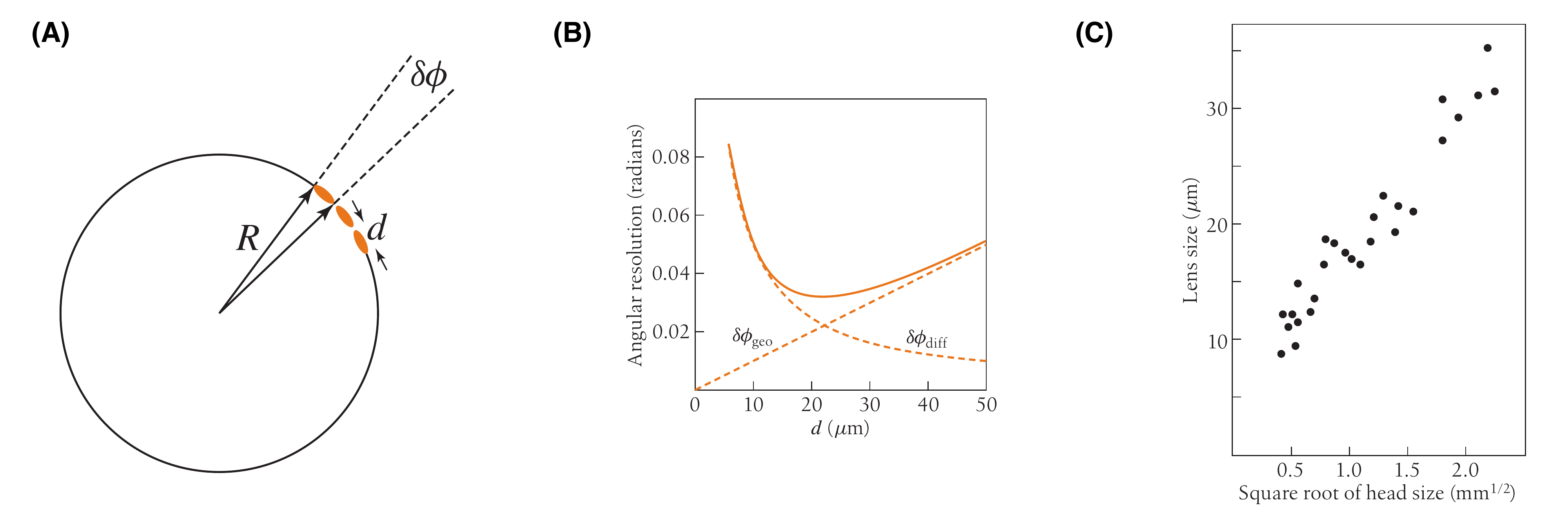}}
\caption{Optics of the insect eye \cite{bialek_2012}. (A) A schematic, in the approximation of a spherical head with radius $R$.  Lenses of size $d$ subtend an angle $\delta\phi = d/R$.  (B) Limits to image resolution from geometry ($\delta\phi_{\rm geo}$) and diffraction ($\delta\phi_{\rm diff}$).  Solid line is a schematic of the combination of limits. (C) The size of lenses in compound eyes as a function of head size, across many species
of insect, redrawn from Barlow \cite{barlow_1952}.\label{insect_eye}}
\end{figure}

The physicist's preference for simple models leads to jokes about spherical cows and horses \cite{devine+cohen_1992}.  Barlow considered the case of an insect with a spherical head (Fig~\ref{insect_eye}A), which is a much better approximation.\footnote{We can do even better and replace the radius of the nominally spherical head with the local radius of curvature of the eye's surface.  There also has been recent progress in going beyond the spherical cow, systematically \cite{lehmann_2025}.}  In the compound eye each receptor cell (roughly) looks out at the world through its own small lens, and in many cases these lenses form a regular lattice; not just the receptors but some layers of processing circuitry also follow this lattice structure.  If lenses of size $d$ are arrayed on a head of radius $R$, then each pixel has an angular size $\delta\phi_{\rm geo} \sim d/R$, and this limits the resolution with which the visual system can form an image.   On the other hand, when lenses become small diffraction blurs the image so that each pixel sees a region of angular size $\delta\phi_{\rm diff}\sim \lambda/d$, where $\lambda$ is the wavelength of light.

As shown in Fig~\ref{insect_eye}B, the geometrical limit pushes for smaller lenses and hence more pixels.  The diffraction limit, on the other hand, pushes for larger lenses.  The compromise is at 
\begin{equation}
\delta\phi_{\rm geo} \sim \delta\phi_{\rm diff} \Rightarrow d^* \sim \sqrt{\lambda R},
\end{equation}
where resolution will be optimized.  Barlow went into the drawers of the natural history museum at Cambridge University to find a variety of insects with varying head sizes, and he verified that the pixel size really does scale with the square root of the head radius, as shown in Fig~\ref{insect_eye}C.  This same argument reappears a decade later in Feynman's lectures on introductory physics, and he compares just with a single species of bee, noting that the lens size $d=d^*$ with $\sim 10\%$ accuracy \cite{feynman}.

This is a very rough theory, but illustrates many important features.  First, at least one piece of the eye's structure is being derived from a physical principle, with no free parameters, and it works reasonably well.  Second, the way we test this idea is by looking across multiple organisms, exploiting rather than avoiding the diversity of organisms.  Finally, this optimization principle carries the hints of its own limitations.  

Maximizing angular resolution makes sense as a design principle for the eye only if other features are not limiting.  As light levels go down, for example, it is advantageous to use larger pixels each of which captures more photons, trading angular resolution for intensity resolution.  With photodetectors that have finite time resolution, acrobatic flight will lead to motion blur and making smaller pixels doesn't help.  These observations suggest that big slow insects flying in the noonday sun should reach the diffraction limit, while others should not, and this is correct in outline.  There have been efforts to make these ideas more precise by asking about the ability of the eye to gather information  rather than form an image \cite{snyder_1977,snyder+al_1977}.    The fact that we have at least a rough theory for the ``design'' of the regular photoreceptor lattice in insect compound eyes raises a question about eyes more like our own, where the arrangement of photoreceptors in much less regular.  We will return to this in \S5, below.

\section{Reasons for doubt}

There are many good reasons to be skeptical of optimization as a basis for theories of living systems.  In some ways the strongest criticism was by Gould and Lewontin \cite{gould+lewontin_1979}, who raised doubt about {\em any} claim that some feature or mechanism serves a particular function.  Their poetic example was an architectural feature in the Cathedral of San Marco, in Venice, the ``spandrels'' that anchor the dome to the rounded arches below.  These spandrels are covered in illustrations that fit beautifully into the biblical narrative of the whole space.  Gould and Lewontin assert that an ``adaptationist'' would view this artistic unity as a cause for the construction of the spandrels, embodying the sentiment that everything we see in living organisms has a function.  But this is backwards, since the artists took advantage of what was necessary to support the structure.  Gould and Lewontin describe this as an ``architectural constraint'' and extrapolate to claim that biological mechanisms are similarly constrained by body plans and their development rather than being necessarily functional.  Their paper has been cited 12,000 times and ``spandrels'' entered the biologists' vocabulary as a pointer to the misattribution of function.

It seems worth pointing out that there no equations and almost no numbers in Gould and Lewontin's discussion.  There certainly is no calculation to support their cartoon of the adaptationist position that spandrels are somehow necessitated for the artistry; indeed it is difficult to imagine what such a calculation would look like.  When we try to articulate optimization principles, as noted above, it is not enough to say that maximizing X is important for the organism.  We need to understand something about what limits this maximization and in some sense whether this is a hard problem.  It is possible, for example, that random mechanisms lead to quite large values for X, in which case maximizing X is unlikely to have much predictive power.  It is not enough to identify a problem that the organism is solving, we need to identify a problem that is hard for the organism to solve.    For the spandrels,  many different shapes would be effective in presenting the artist's vision but only a limited number of shapes solve the hard problem of supporting the dome.  Rather than being an architectural constraint, the spandrels are a good (perhaps even optimal) solution to a different and more challenging problem.  

To give this discussion a more precise mathematical form, it is useful to think about the performance of the organism as a function of the parameters that describe the underlying mechanisms.  
These parameters will take different forms in different examples:  the amino acid sequences of the relevant proteins and the DNA sequences of the enhancer regions that regulate their expression;  the number of copies of each protein that a cell expresses, including for example the number of each type of ion channel in a neuron; the strengths of al the synapses among neurons in a network; and so on.  These parameter spaces typically are very high dimensional---amino acid sequences are long, there are many different kinds of proteins, and many synapses.

It is challenging to build intuition for the behavior of functions that depend on very many parameters.  One possibility is that complex systems with many degrees of freedom can be approximated as random.  This idea goes back at least to Wigner's discussion of energy levels in heavy nuclei, which provided a crucial stimulus for the development of random matrix theory \cite{potters+bouchaud_2020}.  Random pairwise interactions among binary variables defines the energy function of the mean--field spin glass, and the solution of this model provided a prototype for complex systems more generally \cite{mezard+al_1987}.  The energy landscape in this model is rough, with exponentially many local optima organized in a hierarchy and separated by barriers that grow with the number of variables; reasonable local dynamics never really find the true optima, and this connects to ``aging'' and other slow phenomena seen experimentally in real glasses and spin glasses. 

\begin{figure}[t]
\includegraphics[width = \linewidth]{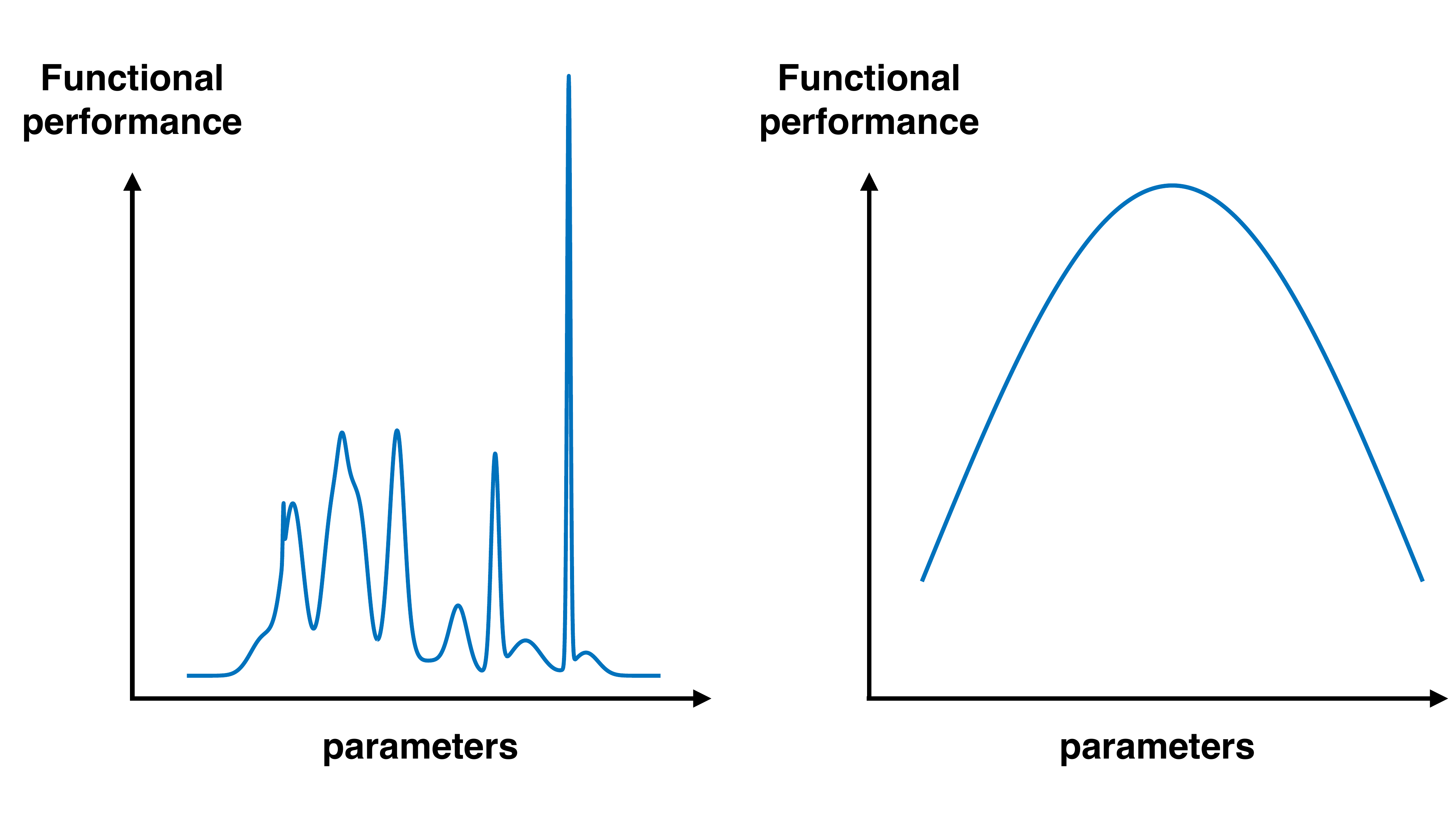}
\caption{Landscapes, the possibility of optimization, and significance of variability.  At left, a schematic ``rough'' landscape in which the true maximum is hard to find and small variations in parameters around this maximum produce catastrophic loss of performance.  At right, the opposite. \label{fig:possible}}
\end{figure}

We can make a one dimensional schematic of a rough landscape, as in the left panel of Fig~\ref{fig:possible}.  Here we are plotting not the energy of a system but the performance at some task, so optima are maxima rather than minima.  We imagine that evolution, learning, and adaptation typically proceed by making small changes in parameters, with guidance or selection from the resulting performance.  In such a rough landscape finding true optima via such local moves obviously is hard, as with the dynamics of the mean--field spin glass.    A related but more subtle point is that even if the system could find the optimum, small changes in parameters would have a large impact on performance.  This is important because we have many examples in which the parameters of biological systems, from the copy numbers of proteins in single cells to the strengths of synapses between neurons, are highly variable \cite{spudich+koshland_76,prinz+al_04,marder+goaillard_06}.  If the landscape is rough then variability is direct evidence against the idea that parameters are tuned to optimal values.

We don't really know what the landscape of performance vs parameters looks like in real biological systems.  It could be simple, as in the schematic in the right panel of Fig~\ref{fig:possible}.  This may be overly optimistic, but it's not impossible.  Obviously if this picture were correct then it would be easy for systems to find their optima, and they could tolerate significant parameter variations.

In summary, the argument of Gould and Lewontin should been seen as a caution not an objection.   As noted at the outset, if we claim that ``maximize X'' is a principle  we should be able to measure X and show that it is close to its optimum.  Any claim that some underlying mechanisms are relevant to maximizing X must be supported by honest calculations.  The feasibility of optimization and the implications of variability depend on  the structure of the landscape, to which we return  explicitly in \S5.

\section{A modern example}

Visual experience is constructed by the brain from   image data collected by the retina.  Artificial intelligence draws attention to complex percepts such as objects and faces, but simpler ones are central to survival and still pose open questions:  arguably there is no case in which we understand quantitatively what algorithm the brain uses to synthesize a percept, how it is implemented in neural hardware, or why it is selected among many possible algorithms.

An accessible example is the visual perception of motion.  For more than a century we have known that spatiotemporal patterns with particular correlation structure elicit the percept of motion even without moving objects \cite{wertheimer_1912}, and for seventy years that the same structures drive insect behavior \cite{reichardt1}; humans and flies even experience analogous illusions \cite{agrochao+al_2020}.  In the fly, a small group of neurons at the output of the visual brain responds to different components of motion \cite{hausen_1982a,hausen_1982b} and drives steering in walking and flying flies \cite{hausen+wehrhahn_1983,hausen_1984}; the receptors act as nearly ideal photon counters over a wide dynamic range \cite{ruyter+laughlin_1996a,ruyter+laughlin_1996b}; and experiments on the intervening layers now culminate in a complete wiring diagram or ``connectome'' in fruit flies \cite{dorkenwald+al_2024}.

When I finished my PhD I didn't know anything about fly vision.  As a postdoc in Groningen, I had the tremendous luck to find myself in the office next door to Rob de Ruyter van Steveninck.  He showed me how he could  make long, stable recordings from motion sensitive neurons at the output of visual computation and also give a fairly complete analysis of the signal and noise properties of the receptors at the input.\footnote{For these kinds of experiments it is conventional to work with larger flies, such as blowflies.  For many other purposes the smaller fruit fly {\em Drosophila melanogaster} is preferred, since in this system we have many genetic tools that allow, for example, targeting fluorescent proteins to particular neurons and classes of neurons.  Differences between species may be important, so one should be careful.}  I didn't know it then but the data rob could collect was on a scale that would not be matched in other systems for a decade.  We saw an opportunity to use this system as a testing ground for a range of theoretical ideas, especially notions of optimality.  This effort continues to the present.

\begin{figure}[t]
\includegraphics[width = \linewidth]{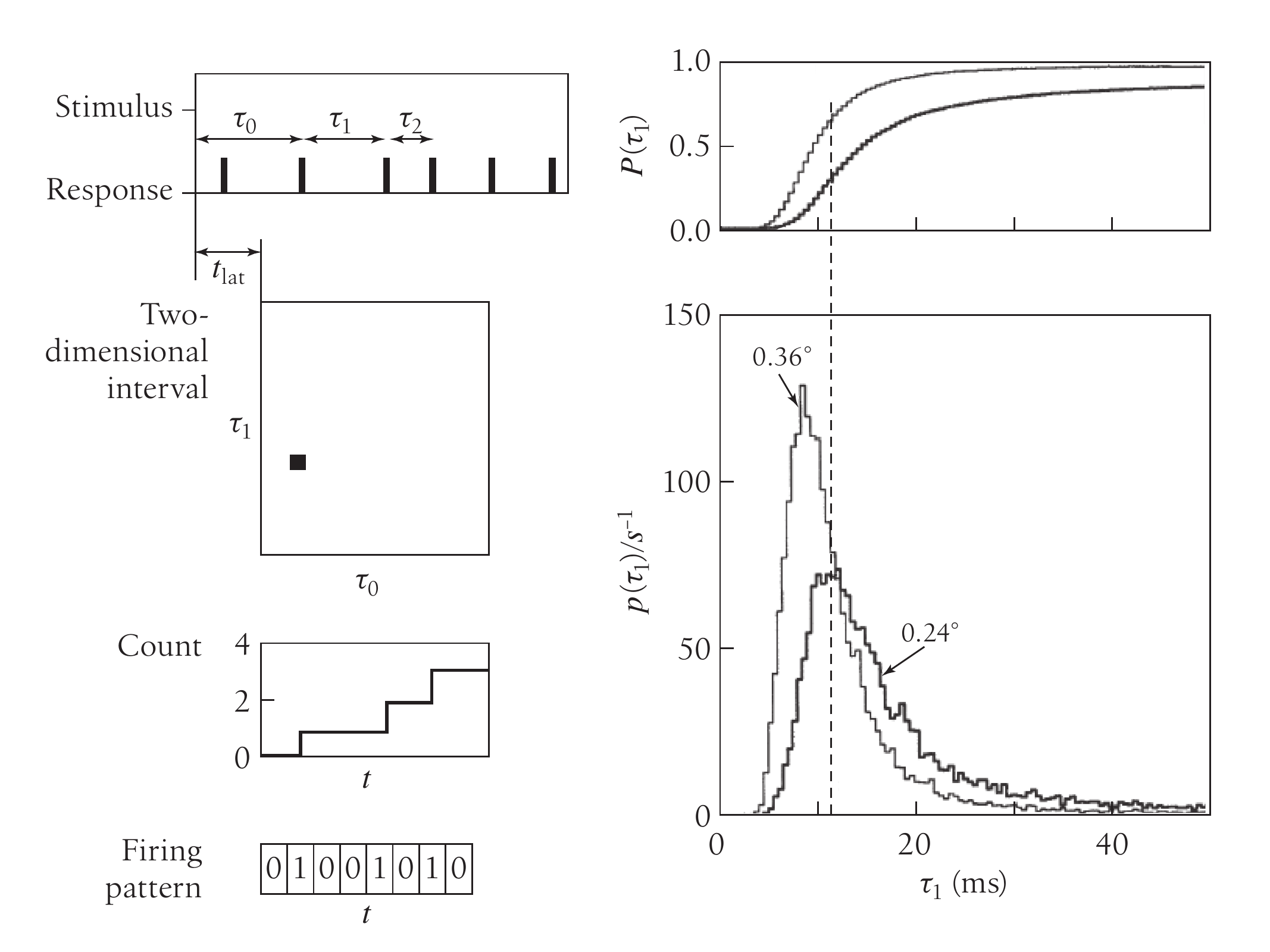}
\caption{Precision of motion estimates encoded by the fly's H1 neuron.  At left, the response to a step displacement isdescribed   by the time until the first spike $\tau_0$ after a latency, the interval from the first spike to the second $\tau_1$, ... .  Alternatively we can  count the spikes, or  discretize time to turn the response into  a binary word.  In each case we can estimate the distribution of responses from many repetitions of the same steps.  At right,  an example with the single interspike interval $\tau_1$. From the cumulative distributions in the top panel we can read off the probabilities of correct identification of each stimulus \cite{ruyter+bialek_1995}.   
\label{fig:H1precision}}
\end{figure}

Among the neurons at the output of the fly's visual brain, H1 is sensitive to wide field motion, and if we show the fly a step displacement it responds with a brief volley of action potentials (Fig~\ref{fig:H1precision}).  These few extra spikes occur in a window of $\sim 40\,{\rm msec}$,  consistent with the behavioral response times of the fly \cite{land+collett_1974}.  We can characterize the response to a step displacement by counting the spikes, by measuring their arrival times relative to the moment of the step, by measuring interspike intervals, or by discretizing time into small bins and thinking of spikes and silences as a binary string.  Because there is some randomness, any of these measures comes from a probability distribution even if the visual stimulus is fixed.  But this distribution shifts when we change the size of the step: large steps produce more spikes with shorter latency and smaller intervals.  These shifts are large enough that we can reach $\sim 75\%$ reliability in distinguishing between steps of $\theta_0 = 0.12^\circ$ and $\theta_1 = 0.24^\circ$ using just one sample of the response, as schematized at right in Fig~\ref{fig:H1precision}A \cite{ruyter+al_1984,ruyter+bialek_1995}.

In this simple step experiment, the precision of motion encoding is $\Delta\theta = |\theta_1-\theta_0| = 0.12^\circ$.  This is an order of magnitude smaller than the angular spacing between photoreceptors on the retina, which is comparable to width of the diffraction blur through the lenses of the compound eye.  Analogous ``hyperacutiy'' phenomena have been known in human vision for more than a century  \cite{westheimer_1981}.  It seems impressive that the fly can detect motion with a precision $10\times$ better than the diffraction limit, but our colleagues who track small particles or even single molecules under the microscope do this routinely.\footnote{For many years hyperacuity seemed mysterious.    It clearly  involves integration over many receptors,  but it was not clear where to find hyperacuity in the response of a single neuron.  Preliminary versions of the results in Fig~\ref{fig:H1precision} were reported in 1984 \cite{ruyter+al_1984}, and this may have been the first such report.  Not only do we see hyperacuity in a single cell, but signals are carried by just a handful of spikes.  I can't remember why it took ten years to get to the final version \cite{ruyter+bialek_1995}.} Should we really be surprised by $\Delta\theta = 0.12^\circ$? What is the physical limit?

Suppose that we look at a pattern of typical contrast $C$ and it moves by an angle $\delta\theta$. A single photodetector element will see a change in contrast of roughly $\delta C\sim  (\delta\theta/\phi_0)$,  where $\phi_0$ is the angular scale of blurring due to diffraction. If we can measure for a time $\tau$ we will count an average number of photons
$R\tau$, with $R$ the counting rate per detector, and shot noise in the photon arrival gives a
fractional precision in intensity of $1/\sqrt{R\tau}$. But fractional intensity is what we mean by
contrast, so  $1/\sqrt{R\tau}$ is   the contrast noise in one photodetector. To get the signal-
to-noise ratio, we should compare the signal and noise in each of the $N_{\rm cells}$ detectors,
then add the squares if we assume (as for photon shot noise) that noise is independent
in each detector while the signal is coherent:
\begin{equation}
SNR \sim N_{\rm cells} \left( {{\delta\theta}\over{\phi_0}}\right)^2 C^2R\tau .
\end{equation}
The experiments in Fig~\ref{fig:H1precision} were done with displays that lead to $R\sim10^4\,{\rm s}^{-1}$ and random spatial patterns that have a contrast $C\sim 0.1$ when viewed through the lenses of the compound eye.  With integration times  $\tau \sim 10-20\,{\rm msec}$, the combination $C^2 R\tau \sim 1$.  H1 integrates over $N_{\rm cells}\sim 10^3$ receptors, so we expect to reach $SNR\sim 1$ at $\delta\theta \sim \phi_0/30 \sim 0.04^\circ$.  This is very rough, but gets within $4\times$ of what we see in the data.    We can do a more realistic calculation using the measured noise properties of the photoreceptors, which deviate from ideal photon counters at higher frequencies, and looking at the evolution of SNR with time after the step and comparing with a more detailed analysis of discriminability in the neural responses; the result is that neural performance is within $2\times$ of the optimum for the first $\sim 30\,{\rm msec}$ of the response.   Rather than asking about discrimination among discrete stimuli we can ``decode'' the sequence of spikes to reconstruct a continuously varying velocity signal, then compare the spectral density of the errors in reconstruction with the limits from photoreceptor noise and diffraction blur; again the neuron is within $2\times$ of the optimum over a reasonable range of frequencies \cite{bialek+al_1991,bialek_1990}.

It would be nice to make a tighter comparison of performance with the optimum.  But we saw in those early data was enough to motivate a theory problem:  What computation does the fly's brain need to do in order to make the best possible estimate of motion?  This problem of optimal estimation is much more general:  we collect sense data and we are interested in estimating some feature of the world.  The key point is that data and features are related only probabilistically, so we can define their joint distribution $P(feature , data)$.  The computation we want is a mapping ${ data} \rightarrow {feature}$ which we can call 
\begin{equation}
feature_{\rm est} = F(data).
\end{equation}
The goal is to get as close as possible to the right answer, and we'll measure this by the mean square error 
\begin{eqnarray}
\chi^2 &\equiv& {\bigg\langle} | feature - feature_{\rm est} |^2{\bigg\rangle}\\
&=& \int D(feature) \int D(data) \, P(feature , data) {\big |} feature - F(data){\big |}^2 .\nonumber\\
&&
\end{eqnarray}
Although very schematic the notation $D(data)$ is meant to remind us that the data typically live in a high dimensional space.

In principle we need to search through all possible functions $F(data)$ to find the one that minimizes $\chi^2$.  This not only seems like a complex problem but also a different problem for every  task.  In fact we can solve the problem formally while staying completely general:
\begin{equation}
\min_{F(data )} \chi^2  \Rightarrow F(data) = \int \int D(feature)\, feature \, P(feature | data) .
\label{condmean}
\end{equation}
In words, the best estimator is the conditional mean.  If we want to be a bit more careful we can think about $data$ in the outside world and $receptor\ signals$ (e.g., photoreceptor voltages) that provide the primary representation of these data at the input to the brain.

For the problem of motion estimation in fly vision, the $data$ are the spatiotemporal patterns of light intensity $I({\vec x},t)$ falling on the retina or the resulting photoreceptor voltages $\{V_{\rm n}(t)\}$, and the feature of interest in the angular velocity $v(t)$.  While direct measurements on the photoreceptors provide a good characterization of the distribution $P[\{V_{\rm n}(t)\} | I({\vec x},t)]$, we know much less about the distribution of the input data $P[v(t); I({\vec x},t)]$.  In early work we made reasonable guesses about the form of this distribution, and were able to make analytic progress in calculating the optimal estimator \cite{potters+bialek_1994}.  To get control over the calculations it was useful to think about different regimes in the overall signal--to--noise ratio of the system.

At low SNR we could write the optimal estimator as a function power series in the receptor cell voltages:
\begin{equation}
v_{\rm est} (t) = \sum_{\rm nm} \int d\tau\int d\tau' \, V_{\rm n}(t-\tau) K_{\rm nm}(\tau, \tau') V_{\rm m}(t-\tau') + \cdots .
\label{vest_1}
\end{equation}
The kernel $K_{\rm nm}$ is dominated by having $\rm n$ and $\rm m$ be neighbors in the lattice of receptors.  The idea of a series expansion for neural computation goes back at least to the work of  Reichardt and Poggio on the analysis of fly visuomotor behavior \cite{reichardt+poggio_1976}.  reaching back even further, in the mid 1950s Hassenstein and Reichardt argued from experiments on beetle behavior that insects detect motion by computing correlations between neighboring pixels in the compound eye \cite{reichardt1}, and this correlator model continues to guide the community's thinking seventy years later \cite{borst+groschner_2023}.  It is attractive to see how the correlator can emerge as one regime of the optimal estimate.

At intermediate SNR the higher order terms represented as $\cdots$ in Eq (\ref{vest_1}) become important, the time scales inside the kernel $K_{\rm nm}$ become shorter, and the restriction that $\rm n$ and $\rm m$ be neighbors becomes tighter.  This crosses over smoothly to the behavior at high SNR, where we can forget about the difference between receptor voltages and light intensities, allowing us to write the estimator directly as a mapping from visual inputs.  Schematically we find
\begin{equation}
v_{\rm est} (t) = - {{\langle (\partial I/\partial t)\cdot (\partial I/\partial x)\rangle}\over{A + \langle  (\partial I/\partial x)^2\rangle}} ,
\label{vest_2}
\end{equation}
where $\partial I/\partial x$ is an approximation to the spatial gradient of the image computed from neighboring points on the photoreceptor lattice, $\langle \cdots \rangle$ denotes an average over all the receptors, and $A$ is a constant that depends on the distribution $P[v(t); I({\vec x},t)]$.

If I can be a little more schematic and not worry about the spatial averaging, then when spatial gradients are large enough the high SNR estimator in Eq~(\ref{vest_2}) becomes
\begin{equation}
v_{\rm est} (t) \sim - {{ (\partial I/\partial t) }\over{  (\partial I/\partial x) }} .
\label{vest_3}
\end{equation}
This ``ratio of gradients'' computation makes sense.  Suppose we have a spatial pattern that moves rigidly at some velocity $v$, and for simplicity we can think just in one dimension so that
\begin{equation}
I(x,t) = I_0(x-vt) \Rightarrow v = - {{ (\partial I/\partial t) }\over{  (\partial I/\partial x) }} .
\end{equation}
So Eq~(\ref{vest_3}) seems like the right answer.  The correlator model, on the other hand, is weird; in particular if we double the image contrast we quadruple the estimated velocity.    You might be willing to accept that flies are weird, but we were trying to derive the optimal estimator and in some regime we found the correlator.  What's going on?  Why isn't the ratio of gradients always the right answer?

Experimentalists know that there are two things you shouldn't do with noisy data.  You shouldn't differentiate, because there usually is more noise than signal at high frequencies so that differentiating lowers the SNR.  You shouldn't divide, because there is a danger of diving by zero and more generally random errors in the denominator become systematic errors in the ratio.  The ratio of gradients  in Eq~(\ref{vest_3}) commits both of these sins, one of them twice.

Optimal estimation always is a compromise between systematic and random errors.  In the absence of noise the ratio of gradients is correct, with zero systematic error (at least in simple settings), but it is {\em very} sensitive to noise.  This not about biology but about math.  Spatial and temporal gradients are correlated in a way that depends on velocity, and if we can trust each gradient completely then we can tae the ratio, which includes all orders of correlations.  But when we add noise the higher order correlations are destroyed first, and if we add enough noise all that remains are the second order correlations, precisely what is picked up in the Reichardt correlator model.

The idea that the (systematically wrong) Reichardt model can emerge from a theory of optimal estimation is very attractive.  Can we take thus take optimization seriously as a theory of the computation that the fly is doing?  The problem is that the correlator emerges as optimal only in some regime.  Is the fly in this regime?  The answer depends on the statistics of the inputs, $P[v(t); I({\vec x},t)]$, and in the initial work we were only guessing at this.  Related ideas have been used to think about motion perception in human and primate vision \cite{simoncelli_1993,weiss+al_2002,stocker+simoncelli_2006,zhang+stocker_2022}, part of a larger project of treating perception as Bayesian inference \cite{knill+richards_2011,mao+stocker_2024}.  All of this work holds out the promise of a principled theory, but if we don't know the distribution of sensory inputs then we are back in the world of parameterzied models.

My colleague Rob de Ruyter van Steveninck set himself the task of building a special purpose camera that captures a ``fly's eye view'' of the world.  He was joined by Shiva Sinha and Charlie Edelson, in two generations of this project \cite{sinha+al_2021,edelson+al_2026}.  The current camera has 187 pixels in a hexagonal array that mimics the compound eye; the spectral sensitivity and optical blur also match those in the fly, and the system samples at $\rm msec$ resolution.  Importantly the camera is bigger than a fly, so it collects orders of magnitude more light, providing an essentially noiseless view of the inputs to the visual system.  The camera is instrumented with motion sensors so that one also has a recording of its angular velocity as we move the camera through a natural environment, e.g. on a walk through the woods.

\begin{figure}[t]
\includegraphics[width=\linewidth]{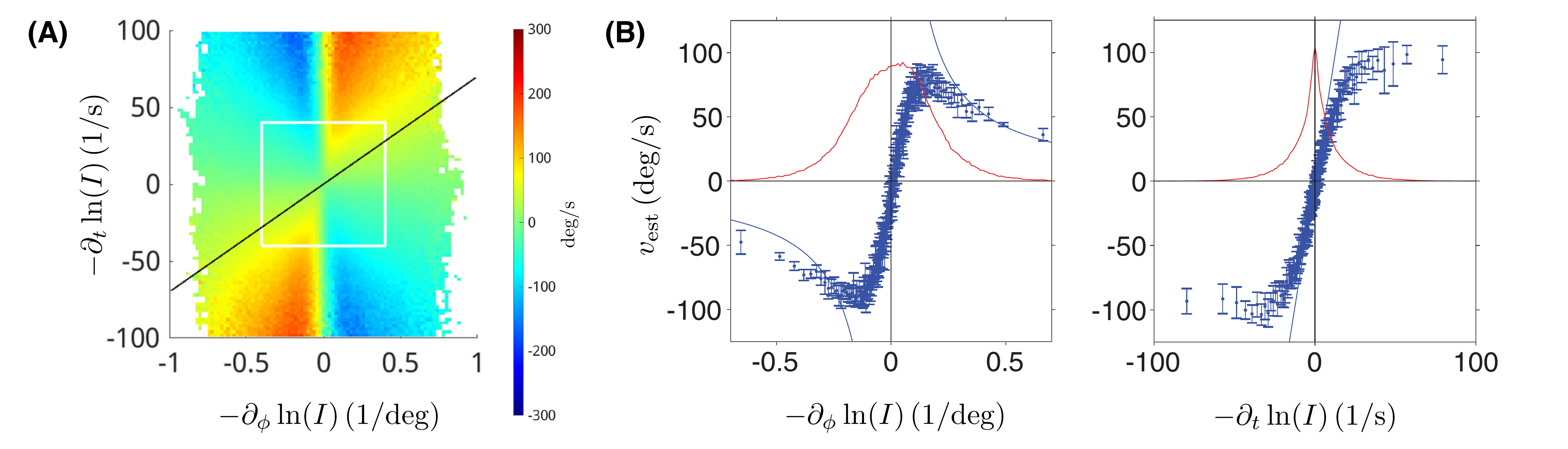}
\caption{Optimal local motion estimation in a natural environment, from Eq~(\ref{local_est}).  (A) Heat map of the estimator \cite{edelson+al_2026}.  The black line is $70\,{\rm deg}/{\rm s}$ in the ratio of gradient computation, and the white square encloses $90\%$ of the data on a walk through the woods. (B) Slices through the optimal estimator at constant temporal  (left: $\partial_t \ln(I) = 21\,{\rm s}^{-1}$)  or spatial (right: $-\partial_\phi \ln(I) = 0.13\,{\rm deg}^{-1}$) derivatives. Blue points are optimal estimates,   error bars are standard deviations across random quarters of an earlier data set \cite{sinha+al_2021}. Blue lines are the predictions of Eq~(\ref{vest_3})  and red lines show the distributions of derivatives along the slices (scaled for clarity).  \label{fig:vest}}
\end{figure}

There is much to be learned just from the problem of {\em local} motion estimation, where we focus on signals from a single pixel and its immediate neighbors.  It is plausible that the only relevant information is contained in the spatial and temporal gradients computed from these signals and perhaps filtered in time.  The data then give us many samples from which we can estimate the distribution $P[v; \partial_\phi \ln (I), \partial_t \ln (I)]$; we focus on the log of the intensity to remove the absolute brightness of the image, which shouldn't matter for the motion estimate.  Again we think in one dimension, corresponding to the horizontal or yaw coordinate $\phi$.  We will also take the raw data, which as noted are essentially noiseless compared with the fly.  Figure~\ref{fig:vest} shows the optimal estimator, making Eq~(\ref{condmean}) explicit:
\begin{equation}
v_{\rm est} = \int dv \, v \, P[v|\partial_\phi \ln (I), \partial_t \ln (I)].
\label{local_est}
\end{equation}

Figure~\ref{fig:vest}A gives a global view of the estimator.  The form is remarkably similar to what we would expect from Eq~(\ref{vest_2}).  At large values of the spatial gradient the contours of constant estimated velocity are straight lines, as with a ratio of gradients computation.  But most of the data that we encounter on this walk in the woods is at smaller gradients, where the contours bend in way that corresponds to the correlator model.  We can see this more explicitly by taking slices through the estimator (Fig~\ref{fig:vest}B).  For the bulk of the data the estimator is linear in both the spatial and temporal gradients, which is what happens if the estimator is proportional to their product.  In the tails of the distribution of spatial gradients we see hints of the ratio of gradients.

The important conclusion is that the correlator computation is optimal under conditions that are relevant in the natural environment.  The results give us a quantitative picture of how the estimator departs from the pure correlator predictions, which should be tested in detail against the responses of motion sensitive neurons.\footnote{As a caution, the system is highly adaptive, so we need to be sure that we are probing the neural response with stimuli drawn from the same distribution that we use in constructing the optimal estimator.  This creates some challenges.} There is much more to be done:
\begin{itemize}
\item We can add noise to the image intensities, simulating the signals in photoreceptors at lower light intensities, then recompute the optimal estimator.  At lower light levels we see longer filtering times and an expansion of the dynamic range over which we see pure correlator behavior, both of which are in agreement with neural responses.
\item Perhaps surprisingly, the optimal estimate of yaw velocity depends on the orthogonal spatial gradient of intensity.  This dependence is not sign specific, and can be interpreted by saying that orthogonal gradients are a source of noise, which again is consistent with neural responses \cite{roy+al_15}.
\item We can divide the walk in the woods into epochs of high and low contrast, and search for adaptation to this longer term statistical structure.
\end{itemize}
Full, quantitative exploration of these ideas still is a work in progress. The results will provide a rich, parameter free theory of what the fly should compute and how this computation should adapt to the statistics of the environment.   The availability of data on these statistics allows us to finally realize the promise of optimization as a complete theory.

\section{The landscape for optimization}

Biological systems seem to be highly variable.  Even in invertebrate neural circuits, where cells can be named and numbered and hence identified across individuals, we see that the selection of ion channels in each cell membrane and the strengths of synapses between cells are highly variable \cite{prinz+al_04,marder+goaillard_06}.  This seems like prima facie evidence against the idea that parameters have been tuned to optimize anything.

\begin{figure}[t]
\includegraphics[width = \linewidth]{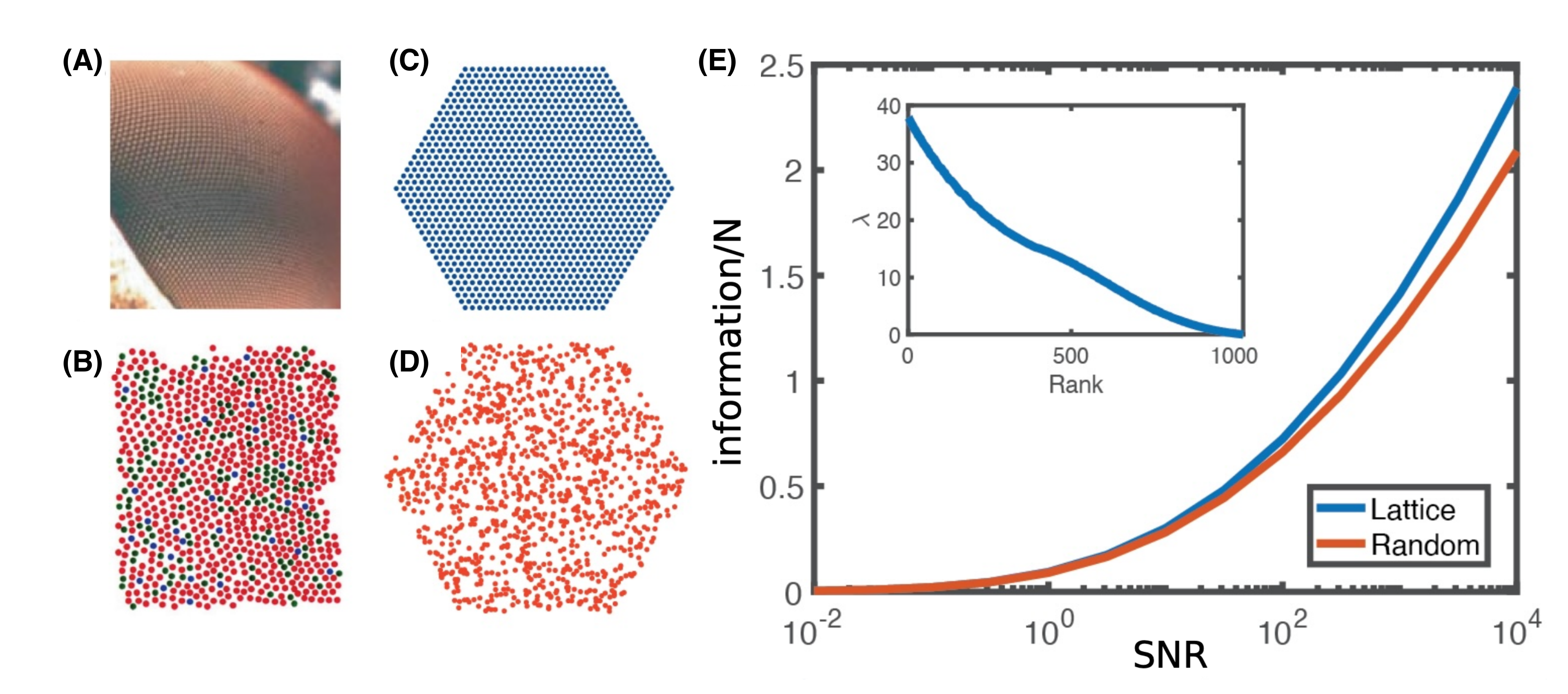}
\caption{Sampling the visual world \cite{bauer+al_2025}. (A) A segment of the fly's eye showing the regular array of lenses, which is echoed by a regular array of receptors and processing circuitry \cite{Leertouwer_1991}. (B) Photoreceptor array in a human retina, with cells colored by their spectral sensitivity  \cite{roorda+al_2001}. (C) A regular lattice of $N=1141$ receptors.  (D) The same number of receptors scattered at random.  (e) Information per cell as a function of the signal--to--noise ratio for the regular (blue) or random (red) arrangements. The range of  $\textit{SNR}$ is chosen to match estimates for human cones in moderate daylight \cite{ruderman+bialek_1994}. Inset shows the spectrum of eigenvalues of the Hessian, from Eqs~(\ref{Hdef}, \ref{Heig}).
\label{eye_order}}
\end{figure}

On the other hand, biological systems can exhibit striking order and reproducibility.  An example is the insect eye that we discussion in \S2:  not only are the lenses the right size to reach the diffraction limit, but in many species these lenses form a very regular array (Fig~\ref{eye_order}A).  The receptor cells and processing circuitry mirror this regular arrangement, and Cajal referred to the insect optic lobes as a ``neural crystal.''  In contrast, vertebrate and especially mammalian neural circuitry is both disordered within individuals and variable across individuals, and this starts in the retina (Fig~\ref{eye_order}A).

Some years ago, at the start of her PhD, Caroline Holmes emphasized to me the problem posed by this contrast between vertebrate and invertebrate brains.  If there were no examples of ordered, reproducible circuitry we might conclude the developmental constraints make such precision impossible (perhaps in the spirit of Gould and Lewontin).  But since order is possible, how can we and other vertebrates ``get away'' with such a mess?  Are we solving different problems?  Does the disorder leave more room for learning a wide variety of tasks?   We were joined by a large group of collaborators to address parallel questions in multiple systems \cite{bauer+al_2025}, but let's start with the photoreceptor array.

 We can imagine an array of receptor cells at positions $\vec x_{\rm n}$.  They respond to a contrast signal $\phi (\vec x )$ that is blurred by a lens and corrupted by added noise,
 \begin{equation}
R_{\rm n} = \int d^2x\, f(\vec x - \vec x_{\rm n}) \phi (\vec x ) + \xi_{\rm n}.
\end{equation}
We'll approximate the blurring function $f(\vec x )$ as Gaussian with a width that matches the mean spacing between receptors, so we have the highest resolution possible without significant aliasing, and take the noise as Gaussian, independent in each cell with variance $\sigma^2$.  There are measurements of the correlation function for the contrast signal in natural environments \cite{ruderman+bialek_1994},
\begin{eqnarray}
\langle \phi (\vec x )\phi (\vec x ' )\rangle &=& \int {{d^2 k}\over{(2\pi)^2}} e^{i{\vec k} \cdot (\vec x - \vec x ')} S (\vec k )\\
 S (\vec k ) &=& {A\over{|{\vec k}|^{2-\eta}}} ,
\end{eqnarray}
and $\eta \sim 0.2$.  If we approximate the fluctuations in $\phi$ as Gaussian then the information that the receptor array captures about the visual world is
\begin{eqnarray}
I\left( \{\vec x_{\rm n}\}\right) &=& \frac{1}{2} \log_2 \det \left(  \mathbf{1}  + \hat C \right)\,  {\rm bits},\\
C_{\rm nm} &=& {1\over {\sigma^2}} \int {{d^2 k}\over{(2\pi)^2}} |\tilde f (\vec k )|^2 S(\vec k ) e^{i{\vec k} \cdot (\vec x_{\rm n} - \vec x_{\rm m})} ,
\end{eqnarray}
where we emphasize that the information depends on the placement of the receptors.
With our assumption of matched Gaussian blur we can choose the mean spacing between receptors as our unit of distance and then the only free combination of parameters is the signal--to--noise ratio in a single cell, $SNR = C_{\rm nn}$.  Results are shown in Fig~\ref{eye_order}E.

As far as we can tell the maximum information transmission always is achieved by a regular array of receptors.  The surprise is that, over a wide range of conditions, a completely random arrangement of receptors is only a few percent worse.   If we consider patterns such as random close packed, this difference is even smaller.  In this case, then, we vertebrates pay hardly any penalty for being less ordered than our invertebrate cousins.  Said another way, the set of possible arrangements that are within some tiny distance of the optimum performance is enormous.

It will be useful to give a local analysis of these deviations from optimality, and to use a slightly more general language.  Let's replace the positions of the receptors by some more abstract set of parameters $\mathbf{\theta} =\{\theta_1, \theta_2, \cdots , \theta_K\}$, and instead of the information we'll write ${\cal F} (\mathbf{\theta})$ for some general measure of functional performance.  There is an optimal setting of the parameters $\mathbf{\theta} = \mathbf{\theta}^*$ where the performance takes its maximum value, and in the neighborhood of this optimum we have
\begin{equation}
{\cal F}(\mathbf{\theta}) = {\cal F}_{\rm max} - {1\over 2}\sum_{{\rm n},{\rm m}=1}^K (\theta_{\rm n} - \theta_{\rm n}^*) H_{\rm nm} (\theta_{\rm m} - \theta_{\rm m}^*)  + \cdots,
\label{Hdef}
\end{equation}
where $\hat H$ is the Hessian matrix.  As usual we can define eigenvectors and eigenvalues,
\begin{eqnarray}
\sum_{{\rm m}=1}^K H_{\rm nm} u_{\mu, {\rm m}} = \lambda_\mu u_{\mu, {\rm n}} .
\label{Heig}
\end{eqnarray}
The eigenvectors described combinations of parameters that have (to leading order) independence effects on the performance $\cal F$, and the eigenvalues describe the strength of these effects.  If we think of the performance as a (negative) energy, then the eigenvalues are the stiffnesses of springs that hold the parameters close to their optimal values.  

In the inset to Figure \ref{eye_order}E we see the Hessian spectrum for the receptor array.  The key result is that there are many very small eigenvalues, so there are directions in parameter space where we can move very far with very little impact on performance.  If we look at the eigenvectors associated with these eigenvalues, they correspond roughly to density waves in the array, with longer wavelengths being connected to smaller eigenvalues, as with soft modes in condensed matter physics \cite{chaikin+lubensky_2000}.

We have explored the dependence of performance on parameters in a wide of examples beyond the photoreceptor array \cite{bauer+al_2025}: readout of position information from transcription factor concentrations in a developing fly embryo, the electrical dynamics of the stomatogastric ganglion in crabs, filtering in recurrent linear neural networks, and image classification in deep (artificial) neural networks.  More recently we have been able to make analytic progress on the landscape for optimization in simple models information transmission and coding \cite{bialek+palmer_2026}.  In each case the measure of functional performance is different and the parameters have different meaning, but we always see the emergence of soft modes.

\begin{figure}[t]
\includegraphics[width=\linewidth]{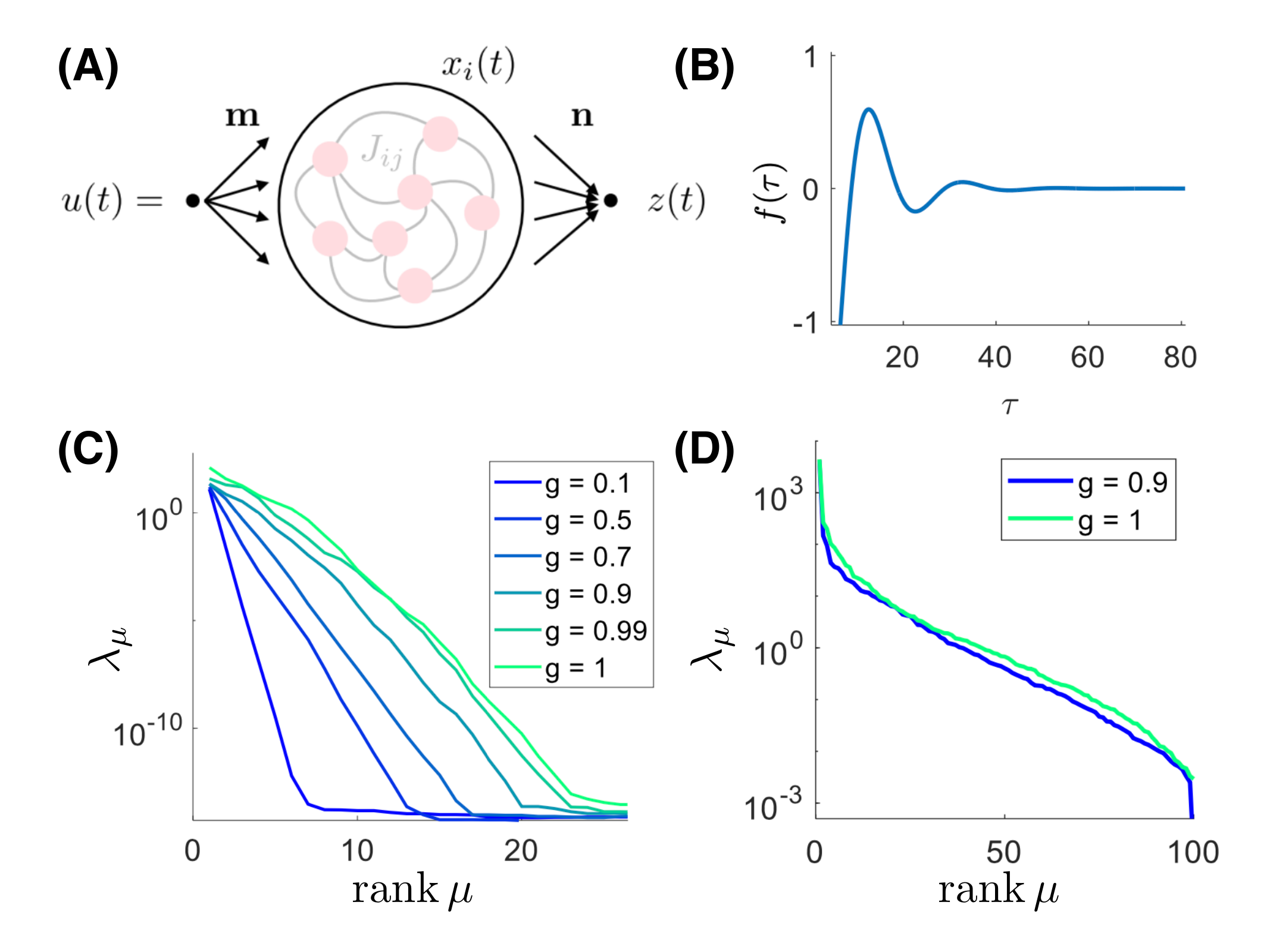}
\caption{A linear recurrent network with $N=100$ neurons \cite{bauer+al_2025}.   (A)  Network architecture.  (B)  The effective filter that relates the input $u(t)$ to the output $z(t)$, as in  Eq~(\ref{filt1}).
    (C) Hessian eigenvalues $\{\lambda_\mu\}$ of the cost function for a random target filter $f_* (\tau )$ when perturbing the readout weights $n_i$; different values of $g$ in colors.
    (D) As in (C) when perturbing the internal weights $J_{\rm ij}$.     
\label{RNN}}
\end{figure}

Since it is an example that is relatively easy to explore on your own, let's take a quick look at the linear recurrent network, as in Fig~\ref{RNN}A.  We have $N$ neurons in which the activity of each neuron is described by a variable $x_{\rm i}$.  This activity relaxes to its resting state on time scale $\tau_0$, and $J_{\rm ij}$ measures the strength of the connection from neuron $\rm j$ to neuron $\rm i$. A single input $u(t)$ that drives all the neurons, each neuron $\rm i$ with a fixed weight $m_{\rm i}$, so the full dynamics are
\begin{equation}
\tau_0 \frac{dx_{\rm i}}{dt} = - x_{\rm i} + \sum_{{\rm j}=1}^N J_{\rm ij} x_{\rm j} + m_{\rm i} u(t) . \label{eq:driven_rnn}
\end{equation}
The output of the network is some linear combination of activities, 
\begin{equation}
z(t) = \sum_{\rm i} n_{\rm i} x_{\rm i}(t) \Rightarrow z(t) = \int dt' f(t-t') u(t'),
\label{filt1}
\end{equation}
for some filter $f(\tau )$ that depends on all the parameters $\mathbf{\theta} \equiv \{J_{\rm ij}; m_{\rm i} ; n_{\rm i}\}$. We measure functional performance as the similarity of this filter to some target,
\begin{equation}
{\cal F}(\mathbf{\theta}) = {\cal F}_{\rm max} - \int d\tau | f(\tau) - f_*(\tau)|^2 .
\label{eq:RNN_cost}
\end{equation}
We can make a generic choice of the target filter $f_*(\tau )$ by setting all the parameters to random values. Note that as we increase the variance of the synaptic strengths, $\langle J_{\rm ij}^2\rangle = g^2/N$, the network has access to a wider range of time scales and these will be reflected in $f_*(\tau )$.  If we vary only  the $N$ readout weights $\{n_{\rm i}\}$  we find the Hessian eigenvalues   in Fig \ref{RNN}B.  Strikingly these eigenvalues are spread over many decades.  At larger $g$, and hence a wider range of time scales,   a larger number of eigenvalues  are significantly nonzero, and as $g\rightarrow 1$ the spectrum becomes almost perfectly uniform on a logarithmic scale.

When we perturb the $N^2$ synaptic strengths $J_{\rm ij}$ we find the eigenvalues Fig~\ref{RNN}C. To understand this result we should think about the eigenvalues and eigenvectors of this matrix.  If perturbations  change only the eigenvectors then we can compensate by adjusting the readout weights.  But then $N(N-1)$ parameters of the synaptic matrix  are truly redundant, so we see only  $N$ significantly nonzero eigenvalues and these are widely distributed (Fig \ref{RNN}C).  They span roughly five orders of magnitude, and except for a few they are distributed uniformly on a logarithmic axis,
\begin{equation}
\lambda_\mu \sim \lambda_1 e^{-\alpha(\mu -1)} ,
\label{sloppy}
\end{equation}
with some slope $\alpha$.

The existence of soft modes in parameter space connects to the pioneering work of Sethna and colleagues on ``sloppy models'' \cite{transtrum+al_15}.  Qualitatively they showed that when fitting multiparameter models to data,
some combinations of parameters are determined precisely, and others are not \cite{brown+sethna_03,gutenkunst+al_07}.  They argued that this is generic \cite{waterfall+al_06}; that the smoothness of models implies that sensitivity to parameters is distributed uniformly on a logarithmic scale \cite{quinn+al_19};   and that this gives a path to model simplification \cite{quinn+al_22}.  

What we have seen is very much in the spirit of sloppy models, but there are two important differences.  First, we are trying to focus not on fitting data but on performance at tasks relevant for the organism's survival.  In place of finding the best fit we are searching for optimal function.  Second, the parameters are not constructs of some model, but rather physical properties of the underlying mechanisms---the affinity of a transcription factor for its binding sites along the genome, the number copies of different proteins in a cell, the positions of cells in the retina, the strengths of synaptic connections between neurons.  There is no sense in which we can simplify the system by projecting out directions in parameter space that have little effect on the performance, not least because these are combinations of parameters.  If there are soft modes in the parameter space of a genetic network, for example, this does not mean, for example, that there is an equivalent network with fewer genes.  The observation of a sloppy spectrum does mean that systems with widely varying parameters can be very close to optimal.  

To make these idea precise let's imagine a population of cells, networks, or organisms that choose their underlying parameters $\mathbf{\theta}$ at random from a distribution $P(\mathbf{\theta})$.  This distribution supports an average performance
\begin{equation}
\bar{\cal F} = \int d^K\theta\, P(\mathbf{\theta}) {\cal F}(\mathbf{\theta}) .
\label{barC_fix}
\end{equation}
In the absence of any other constraints we expect that the parameters will be as variable as possible while obeying the mean value in Eq (\ref{barC_fix}).  To make the distribution as variable as possible we maximize its entropy at fixed $\bar{\cal F}$, and the result is a Boltzmann distribution
\begin{equation}
P(\mathbf{\theta})  = \frac{1}{Z(T)} \exp\left[ + \frac{1}{T} {\cal F}(\mathbf{\theta})\right],
\label{boltz}
\end{equation}
where the ``temperature'' $T$ must be set so that Eq (\ref{barC_fix}) is satisfied and $Z(T)$ enforces the normalization of the distribution.    We can also think of this Boltzmann distribution in several other ways: as a compromise between performance and the information cost of controlling parameters,\footnote{Thanks to Mirnada Holmes--Cerforn for this perspective.} as an approximation to the distributions that emerge in evolutionary dynamics \cite{bodova+al_2016,bodova+al_2021},\footnote{This idea, expressed more intuitively, was implicit in the earliest discussions of the ensemble of DNA sequences that form binding sites for particular proteins \cite{hippel+berg_1986,berg+hippel_1987,berg+hippel_1988}.} as a generalization of maximal growth rates in flux balance analysis of bacterial metabolism \cite{demartino+al_2017,demartino+al_2018}, or as a Bayesian prior on the space of parameters to be used in fitting data \cite{mlynarski+al_2021}. Here we focus on its implications.

If we can neglect higher order terms in Eq~(\ref{Hdef}) then the distribution of parameters is like the equilibrium statistical mechanics of particles held to the origin by springs, and we can immediately calculate the mean ``energy'' $\bar{\cal F}$  and the entropy:
\begin{eqnarray}
\bar{\cal F} &=& {\cal F}_{\rm max} - {K\over 2} T\\
S &=& {1\over 2}\sum_{\mu =1}^K \ln(2\pi e T/\lambda_\mu ).
\end{eqnarray}
We can put these together to give
\begin{equation}
\bar{\cal F} = {\cal F}_{\rm max} -{{K\Lambda}\over{4\pi e}} e^{S/K},
\label{Fresult}
\end{equation}
where $\Lambda$ is the geometric mean of the Hessian eigenvalues,
\begin{equation}
\Lambda = \left( \prod_{\mu = 1}^K \lambda_\mu\right)^{1/K}.
\end{equation}

The ``functional ensemble'' that we have defined by $P(\mathbf{\theta})$ is a statistical mechanics problem with $K$ degrees of freedom.  We expect that the entropy will be extensive so it is reassuring that it is the entropy per parameter, $S/K$, that enters Eq~(\ref{Fresult}).  We also expect that energy will be extensive, and correspondingly the deviation of $\cal F$ from its maximum value also is proportional to $K$.  But the coefficient of proportionality is $\Lambda$.  If we have the ``exactly sloppy'' eigenvalue spectrum from Eq~(\ref{sloppy}) then
\begin{equation}
\Lambda = \lambda_1 e^{-\alpha(K-1)/2}.
\end{equation}
This vanishes exponentially as the number of parameters becomes large, so that $K\Lambda \rightarrow 0$ as $K\rightarrow\infty$.  This means that when we have many parameters with a sloppy spectrum we can have an ensemble of organisms with finite entropy per parameter yet the average performance is abitrarily close to optimal. Variability and optimality coexist.

There is more to say, some of which is highlighted in Ref \cite{bauer+al_2025}:
\begin{itemize}
\item Our main result is more general than for the exactly sloppy spectrum.  As an example, if we have the slower decay $\lambda_\mu \sim \mu^{-p}$, then $K\Lambda \rightarrow 0$ at large $K$ if $\mu > 1$.
\item This seems not to be a generic consequence of being in high dimensional parameter spaces.  As an example, if we train a deep network to classify images with (unnatural) randomized labels then the Hessian spectrum fails to be sloppy.
\item If learning or evolution corresponds to something like Brownian motion on the performance landscape, then there is an entropic force that favors optima with have soft modes \cite{dicarlo+al_2025}.
\item There is considerable interest in the Hessian spectrum as a more general diagnostic in machine learning problems, with hints that soft modes are connected to generalization ability.  For a recent view see \cite{bernstein+al_2026}.
\item We have discussed continuous parameterizations but one can find analogs in discrete spaces.  Thus many amino acid sequences map into the same protein structure \cite{li+al_96,li+al_98} and many different sequential decision strategies provide optimal solutions to foraging problems \cite{ma2024vast}. 
\end{itemize}
The summary is that, contrary to a widely held intuition, near--optimal performance  does not require fine tuning of parameters.  Importantly, this tolerance for parameter variation is not a retreat from optimality; rather it is a feature of the landscape for optimization itself.   Optimization principles  thus {\em predict} that there should be substantial variation in the underlying parameters.

\section{Coda}

More than a century after the first hints that we can count single photons, optimization principles continue to be productive.    I am especially excited that ideas which started in the context of neural coding are so fruitful in thinking about the genetic networks that control early embryonic development in the fruit fly \cite{bialek_2024}; connections between theory and experiment in this system are especially detailed, quantitative, and parameter free. 

The optimization problems  relevant for biological systems are challenging.  Often we solve much simpler problems---linear receptive fields,  coding in populations of highly idealized neurons, ... .  Again I think recent work on genetic networks provides an example of how more realistic calculations make closer contact with experiment \cite{sokolowski+al_2025}. In the fly motion estimation problem (\S4), new data on the statistics of the input to the computation are driving a more detailed analysis with the goal of predicting quantitative features of the optimal estimator and its adaptation that can be compared with neural responses.  Importantly the predictions are parameter free.

Exploration of the landscape for optimization is just starting.  There are precursors in experiments on protein evolution \cite{carneiro+hartl_10,papkou+al_23} and  connections to machine learning.  Indeed the success of modern machine learning is dramatic evidence that optimization in high dimensional parameter spaces {\em works}, despite many reasons why it should not.  These successes  should embolden us to revisit our ambitions for theories of natural systems.

\section*{Acknowledgments}

The ideas in this lecture have been shaped by 40+ years of conversation, collaboration, and friendship with Rob de Ruyter van Steveninck.  In addition to Rob,  examples I discussed are based on work with M Bauer, CJ Edelson,  CW Goddard, CM Holmes, K Krishnamurthy,  SE Palmer, R Pang, M Potters, S Setayeshgar,  DJ Schwab,  SR Sinha, and L Susman.  This work was supported in part by the US National Science Foundation and the Simons Foundation.  Special thanks to Antonio Roque who has brought  insight, energy, and warmth to the   LASCON series, and to the students for making it all so much fun.

\end{document}